\documentclass[acmsmall,screen]{acmart}

\AtBeginDocument{%
  }

\acmPrice{15.00}
\acmISBN{978-1-4503-XXXX-X/18/06}

\usepackage{amsmath,amsfonts}
\usepackage{graphicx}
\usepackage{textcomp}
\usepackage{xcolor}
\usepackage{listings}
\usepackage{enumitem}
\usepackage{microtype}
\usepackage{caption}
\usepackage{subcaption}
\usepackage{booktabs}
\usepackage[ruled,vlined,linesnumbered]{algorithm2e}
\usepackage{setspace}
\usepackage[most]{tcolorbox}
\usepackage{bbding}
\usepackage{multirow}
\usepackage{multicol}
\usepackage{colortbl}

\usepackage{xspace}
\newcommand{\dataset}{FAUN-Eval\xspace}
\definecolor{codegreen}{rgb}{0,0.6,0}
\definecolor{codegray}{rgb}{0.5,0.5,0.5}
\definecolor{codepurple}{rgb}{0.58,0,0.82}
\definecolor{codeblue}{rgb}{0.0, 0.28, 0.67}
\definecolor{bg}{rgb}{0.95,0.95,0.95}
\lstdefinestyle{PythonStyle}{
        language=Python,
        morekeywords={as},
        keywordstyle=\color{magenta},
        stringstyle=\color{codeblue},
        commentstyle=\color{codegreen},
        numberstyle=\tiny\color{codegray},
        extendedchars=false, 
	basicstyle=\ttfamily\scriptsize,
        numbers = left,
        backgroundcolor=\color{bg},
        captionpos=b,
        frame = shadowbox, 
	breaklines=true,
}
\begin{document}

\title{A Real-World Benchmark for Evaluating Fine-Grained Issue Solving Capabilities of Large Language Models}

\author{Ruida Hu}
\authornote{Work done during an internship at ByteDance.}
\email{200111107@stu.hit.edu.cn}
\affiliation{
    \institution{Haribin Institute of Technology, Shenzhen}
    \city{Shenzhen}
    \country{China}
}

\author{Chao Peng}
\authornote{Corresponding authors.}
\email{pengchao.x@bytedance.com}
\affiliation{
    \institution{ByteDance}
    \city{Beijing}
    \country{China}
}

\author{Jingyi Ren}
\email{jingyi.422@bytedance.com}
\affiliation{
    \institution{ByteDance}
    \city{Shenzhen}
    \country{China}
}

\author{Bo Jiang}
\email{jiangbo.jacob@bytedance.com}
\affiliation{
    \institution{ByteDance}
    \city{Shenzhen}
    \country{China}
}

\author{Xiangxin Meng}
\email{mengxiangxin.1219@bytedance.com}
\affiliation{
    \institution{ByteDance}
    \city{Beijing}
    \country{China}
}

\author{Qinyun Wu}
\email{wuqinyun@bytedance.com}
\affiliation{
    \institution{ByteDance}
    \city{Beijing}
    \country{China}
}

\author{Pengfei Gao}
\email{gaopengfei.se@bytedance.com}
\affiliation{
    \institution{ByteDance}
    \city{Beijing}
    \country{China}
}

\author{Xinchen Wang}
\authornotemark[1]
\email{200111115@stu.hit.edu.cn}
\affiliation{
    \institution{Haribin Institute of Technology, Shenzhen}
    \city{Shenzhen}
    \country{China}
}

\author{Cuiyun Gao}
\authornotemark[2]
\email{gaocuiyun@hit.edu.cn}
\affiliation{
    \institution{Haribin Institute of Technology, Shenzhen}
    \city{Shenzhen}
    \country{China}
}


\begin{abstract}
Automatically resolving software issues
is crucial for software development in practice, impacting the software quality and user experience.
The process of resolving real-world issues encompasses tasks such as question-answering (QA), fault localization, and code editing.
Existing benchmarks such as HumanEval fall short in their ability to assess LLMs’ proficiency in solving
issues within a codebase. 
Although benchmarks like SWE-Bench are designed to evaluate the LLMs’ capability to handle real-world GitHub issues, 
the end-to-end evaluation method cannot provide granular insights on the performance of subtasks involved in issue solving.

To address existing deficiencies in benchmarking LLMs for practical software engineering tasks, we introduce \textbf{\dataset}, a benchmark specifically designed to evaluate the \textbf{F}ine-gr\textbf{A}ined iss\textbf{U}e solvi\textbf{N}g capabilities of LLMs.
\dataset systematically assesses LLMs across three distinct tasks: QA, fault localization, and code editing. This benchmark is constructed using a dataset curated from 30 well-known GitHub repositories. For each entry, issue and pull request (PR) pairs are meticulously compiled and validated using cross-referencing and keyword verification methods.
\dataset includes 300 entries and employs both LLM and manual checks to ensure data quality.
We evaluate ten LLMs with \dataset, including four closed-source and six open-source models.
Our experimental results reveal several key findings. We find that the top-performing LLMs differ across the different tasks. Additionally, features in issues may lead LLMs to generate incorrect information. Moreover, models may vary in their proficiency with texts of different lengths.
    
\end{abstract}



\maketitle

\section{INTRODUCTION}
Large language models (LLMs) are increasingly being integrated into tools such as chatbots and coding assistants, showcasing their potential to transform various software engineering tasks~\cite{hou2023large}.
As a result, the research community has begun exploring how LLMs can be further leveraged to assist with more complex real-world tasks encountered in software development~\cite{fan2023automated, macneil2023experiences,ross2023programmer, yang2024swe, zhang2024autocoderover, liu2024large}.
SWE-agent~\cite{yang2024swe}, for example, designs a custom agent-computer interface (ACI) that allows LLM to act as agents to interact with the repository environment through actions such as reading, editing files and running bash commands to resolve GitHub issues and develop features.
Numerous LLM-based approaches~\cite{zhang2024autocoderover, tao2024magis, xia2024agentless, ma2024understand, liu2024marscode} have been proposed to identify locations that need modification and propose code edits to solve these issues.

Using LLMs to solve real-world issues usually consists of three phases: problem analysis, fault localization, and code edit (patch) generation.
Problem analysis involves understanding the issue reported by a developer or user, often requiring the model to interpret the natural language description of the problem and propose solutions of the problems.
Fault localization requires the model to pinpoint the specific file or section of code responsible for the reported issue, which is particularly challenging in large and complex codebases.
Once the faulty code file is identified, the final phase, code edit generation, involves creating the appropriate changes to resolve the issue.


SWE-Bench is a prominent benchmark designed to evaluate LLMs on issue solving in an end-to-end manner by requiring them to generate patches and pass the corresponding unit tests.
This benchmark addresses the limitations of existing coding benchmarks such as HumanEval~\cite{chen2021evaluating} by presenting tasks that require models to understand and coordinate changes in large codebases involving multiple functions and files.
Although existing benchmarks provide an important step forward by evaluating models in a realistic development context, they still face two major challenges:

\begin{enumerate}
    \item \textbf{Inadequate Handling of Complex Codebases.} Existing LLM benchmarks primarily focus on relatively small, self-contained problems. These benchmarks do not fully capture the complexity of real-world software maintenance tasks, which involve navigating large codebases with interconnected components. Consequently, LLMs struggle with tasks requiring multi-file changes and deeper understanding of the software architecture.
    \item \textbf{Limited Fine-Grained Evaluation.} Benchmarks such as SWE-bench evaluate LLMs in an end-to-end manner but fail to provide insights into the performance of LLMs on individual subtasks, such as problem analysis, fault localization, and code edits. The lack of granular evaluation metrics makes it difficult to pinpoint where current models fall short and how they can be optimized for specific aspects of issue resolution.
\end{enumerate}

To address these challenges, we present an evaluation of large language models in solving GitHub issues across three core tasks: code question answering, fault localization, and code editing.
These tasks represent common problem-solving scenarios encountered in software maintenance.
In code question answering, the model is tasked with providing a response to issue descriptions, which reflects a typical response from the repository maintainer.
For fault localization, the goal is to identify the faulty file that needs modification.
Lastly, the code edit task involves generating the appropriate code changes based on issue and pull request information.

To evaluate the performance of LLMs on these tasks, we conduct extensive experiments with state-of-the-art models, including GPT, DeepSeek Coder and Gemini series.
For each task, we assess the accuracy of the model predictions against ground truth, using metrics such as Exact Match (EM), BLEU, CodeBLEU, ROUGE-L, ROUGE-1, and Edit Similarity.
We find that different models excel in specific areas, such as Gemini-1.5-Pro and GPT-4o for fault localization, and DeepSeek-Coder-V2 for code editing, underscoring the need for task-specific optimization.
Proprietary models do not consistently outperform open-source ones, nor do larger models guarantee superior results, with open-source models like DeepSeek-Coder-V2 often performing better in certain tasks.
Furthermore, LLMs struggle with real-world question answering, where their generated responses frequently lack accuracy, and some models fail to follow task instructions, particularly in fault localization.
These findings suggest that model selection should be tailored to the challenges of each task and improvements are needed to enhance the practical performance of LLMs in software engineering tasks.

In sum, the paper makes the following contributions.

\begin{enumerate}
    \item \textbf{Task-Specific Evaluation Framework.} We introduce a novel evaluation framework that breaks down the issue-solving process into three core subtasks: code question answering, fault localization, and code editing. This fine-grained approach allows for a more detailed evaluation of LLM performance across each subtask, providing clearer insights into model strengths and weaknesses.
    \item \textbf{Real-World Dataset from GitHub Issues.} We build a comprehensive dataset derived from real-world GitHub issues and corresponding pull requests. This dataset encompasses a wide variety of tasks that reflect the complexities of actual software maintenance, offering a more realistic assessment of LLM capabilities.
    \item \textbf{Comprehensive Experimental Results.} We evaluate state-of-the-art LLMs across the subtasks of issue resolution. Our experiments reveal key insights into the effectiveness of these models, highlighting areas where they perform well and where improvements are needed.
    \item \textbf{Guidance for Model Selection and Optimization.} By providing detailed performance metrics for each subtask, our evaluation framework helps guide better model selection and optimization strategies. This allows developers and researchers to make informed decisions about which models to deploy in specific software engineering tasks and how to fine-tune them for higher performance in real-world scenarios.
\end{enumerate}

\section{BACKGROUND AND RELATED WORK}
\subsection{LLMs and LLM-based Agents}

LLMs are particularly effective in tasks that treat software engineering challenges as text, code, or data analysis problems, owing to their ability to process large-scale data, understand programming languages, and generate useful software artifacts such as code snippets, summaries, and bug fixes.
For example, models such as Codex~\cite{chen2021evaluating} and GPT-4 have demonstrated impressive abilities in solving programming problems with high accuracy, including code summarization~\cite{geng2024large, ahmed2024automatic}, bug detection and localisation~\cite{du2024generalization, qin2024agentfl}, software testing~\cite{sun2023smt, yuan2024evaluating, lemieux2023codamosa}, root cause analysis~\cite{chen2024automatic} and program repair~\cite{lin2024one, wei2023copiloting}. These applications leverage the strength of LLMs in understanding both natural language and code, enabling them to contribute significantly to the automation and enhancement of software development processes.

To further extend the capabilities of standalone LLMs and enable them to tackle more complicated and collaborative tasks, researchers have also explored LLM-based agents by integrating external tools, planning mechanisms, and multi-agent collaborations~\cite{liu2024large}.
It is allowed for agents to break down complex SE tasks into manageable subtasks via planning.
For example, agents can be assigned specialized roles such as fault localization, code generation, or patch refinement, and collaborate to achieve a common goal.
LLM-based agents have been successfully applied to various SE tasks, including code generation~\cite{olausson2023self}, static analysis~\cite{hu2023large}, testing~\cite{yuan2024evaluating}, and debugging~\cite{qin2024agentfl}.

\subsection{Existing Benchmarks for Issue Resolving}

\definecolor{darkgreen}{rgb}{0,0.5,0}
\newcommand{\cross}{\textcolor{red}{\textbf{\XSolidBrush}}}
\newcommand{\tick}{\textcolor{darkgreen}{\Checkmark}}

\begin{table}[!htbp]
    \caption{The comparison between existing benchmarks and \dataset. DCM and PQC refer to Dataset Construction Method and Pair Quality Confirmation, respectively. MCQ stands for Multiple-Choice Questions. AC, DG, MR, and FF denote Automated Crawling, Data Generation, Manual Review, and Feature Filtering, respectively. "-" indicates unknown, and FL means fault localization.}
    \fontsize{7.2}{11}\selectfont
    \centering
    \setlength{\tabcolsep}{0.8mm}
    \begin{tabular}{c|c|ccc|c|c|c|c|c}
    \toprule
    \textbf{Dataset} & \textbf{Format} &\textbf{\#Langs} & \textbf{\#Repos} & \textbf{\#Samples} & \textbf{DCM} & \textbf{PQC} & \textbf{Real-world} & \textbf{Issue-solving} & \textbf{Tasks}
\\
\midrule
MMLU~\cite{hendryckstest2021}
& MCQ
& - 
& - 
& 15,908  
& AC
& \cross
& \tick
& \cross
& code QA
\\

CodeQA~\cite{liu2021codeqa}
& QA pair
& 2
& -
& 190,000
& AC, DG
& \cross
& \cross
& \cross
& code QA
\\

CS1QA~\cite{lee2022cs1qa}
& QA pair
& 1
& - 
& 9,237
& AC, MR
& \tick
& \tick
& \cross
& code QA
\\

CodeApex~\cite{fu2023codeapex}
& MCQ
& -
& -
& 250
& MR
& \tick
& \tick
& \cross
& code QA
\\

\cellcolor{blue!10}\textbf{\dataset-QA}
& \cellcolor{blue!10}{QA pair}
& \cellcolor{blue!10}{5}
& \cellcolor{blue!10}{30}
& \cellcolor{blue!10}{300}
& \cellcolor{blue!10}{AC, FF, MR}
& \cellcolor{blue!10}{\tick}
& \cellcolor{blue!10}{\tick}
& \cellcolor{blue!10}{\tick}
& \cellcolor{blue!10}{code QA}
\\ 
\midrule

Defects4J~\cite{just2014defects4j}
& code pair
& 1
& 17
& 835
& MR
& \tick
& \tick
& \cross
& code repair
\\

TFix~\cite{berabi2021tfix}
& code pair
& 1
& -
& \textasciitilde105,000
& AC
& \cross
& \tick
& \cross
& code repair\\

FixJS~\cite{csuvik2022fixjs}
& code pair
& 1
& -
& \textasciitilde324,000
& AC
& \cross
& \tick
& \cross
& code repair\\

TypeBugs~\cite{oh2022pyter}
& code pair
& 1
& 15
& 93
& AC, MR
& \cross
& \tick
& \cross
& FL, code repair\\

\cellcolor{blue!10}\textbf{\dataset-fix}
& \cellcolor{blue!10}{code pair, fix pair}
& \cellcolor{blue!10}{4}
& \cellcolor{blue!10}{17}
& \cellcolor{blue!10}{300}
& \cellcolor{blue!10}{AC, FF, MR}
& \cellcolor{blue!10}{\tick}
& \cellcolor{blue!10}{\tick}
& \cellcolor{blue!10}{\tick}
& \cellcolor{blue!10}{FL, code editing}
\\

    \bottomrule
    \end{tabular}
    \label{tab:data_compare}
\end{table}

Various benchmarks have been designed to evaluate LLMs in specific domains. For evaluating end-to-end software development, benchmarks such as HumanEval~\cite{chen2021evaluating}, APPS~\cite{hendrycks2021measuring} and MBPP~\cite{austin2021program} have been widely used, but these focus on small-scale code completion and generation tasks.

\subsubsection{Task-specific benchmarks}

Question-answering benchmarks, such as MMLU~\cite{hendryckstest2021},
evaluate models in zero-shot and few-shot settings and assess their compability in answering question across 57 subjects across science, technology, engineering and mathematics. CodeQA~\cite{liu2021codeqa} contains a dataset of 119,778 Java and 70,085 Python code-question pairs. CS1QA~\cite{lee2022cs1qa} contains 9,237 question-answer pairs extracted from chat logs from introductory Python classes. CodeApex~\cite{fu2023codeapex} evaluate LLMs on C++ code generation and correction.
However, these benchmarks do not target questions to solve issues but focus on answering general code-related questions, such as programming language syntax and programming basics.

To evaluate bug fixing capabilities, Defects4J~\cite{just2014defects4j}, TFix~\cite{berabi2021tfix}, FixJS~\cite{csuvik2022fixjs}, TypeBugs~\cite{oh2022pyter} are notable benchmarks for Java, JavaScript and Python programs.
However, they are designed for end-to-end program repair and do not use text input for fault localization and repair, which is distinct from real-world issue-solving scenarios.

\subsubsection{End-to-end Benchmarks}

SWE-Bench~\cite{jimenez2023swe} is a benchmark developed to assess LLMs on complex, real-world software engineering tasks derived from GitHub issues and corresponding pull requests across 12 widely used Python repositories.
SWE-Bench presents tasks that require models to come up with patches to solve the issue.
However, SWE-Bench is not well-suited for evaluating the individual subtasks involved, as even top-performing models, achieve a success rate of only 4.33\% and meaningful comparisons between different models are difficult to make.
This limitation highlights a critical gap: SWE-Bench provides results on end-to-end issue solving but offers little guidance on how LLMs perform on the crucial subtasks that form the foundation of issue resolution.

As summarized in Table~\ref{tab:data_compare}, despite existing efforts on LLM evaluation, there remains a gap in benchmarks designed to evaluate subtasks within software issue resolution workflows.
This has motivated the development of our work, which aims to better understand how LLMs perform on these critical subtasks, ultimately guiding more effective model selection and optimization strategies.



\section{APPROACH}
\begin{figure}[t]
	\centering
	\includegraphics[width=1\textwidth]{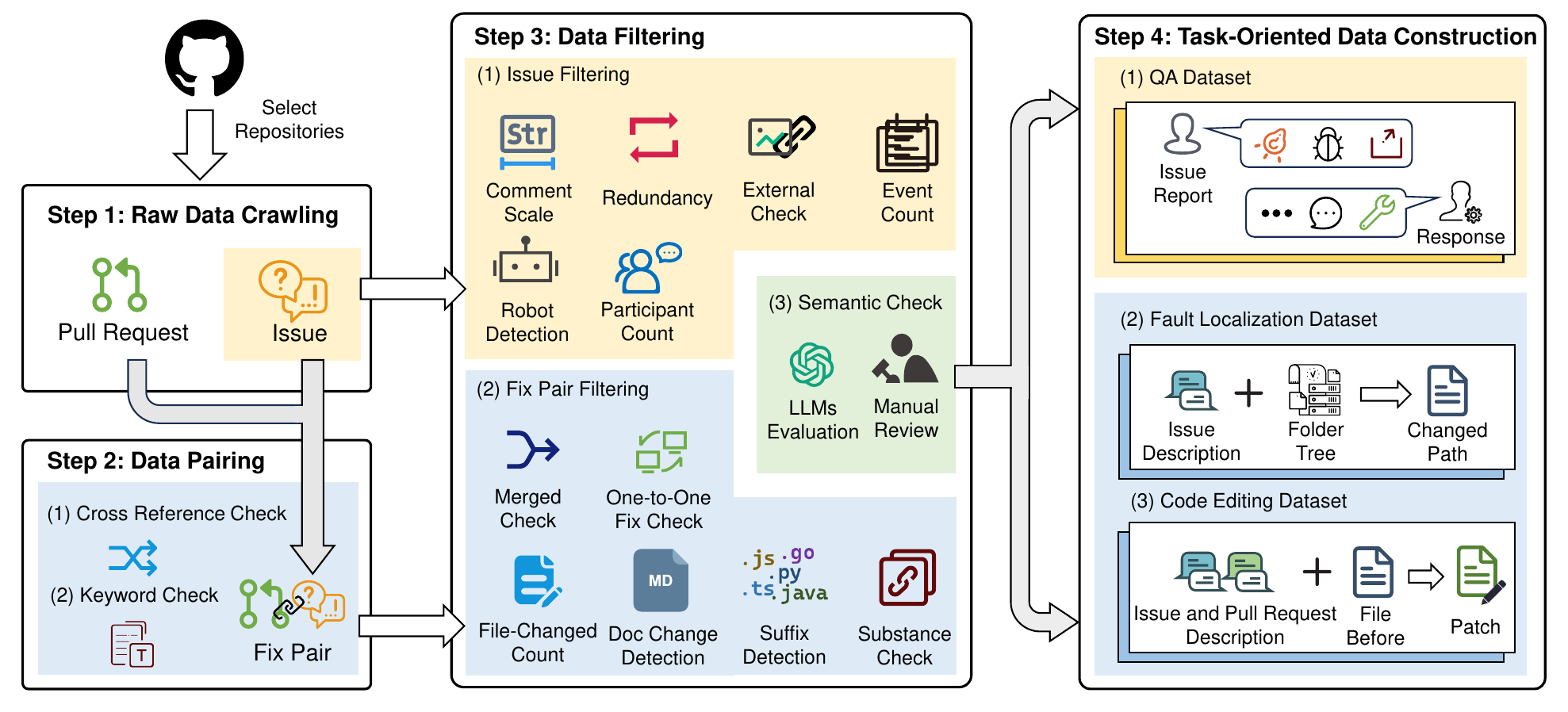}
    \caption{The architecture of our data collection framework.}
\label{fig:architecture}
\end{figure}

The architecture of \dataset is shown in Figure \ref{fig:architecture}. We start with the raw data crawling, and we extract metadata related to issues and Pull Requests (PRs) from GitHub. Following this initial data collection, the process continues through three key steps to produce the final dataset: (1) The data pairing step, where issues are matched with corresponding Pull Requests that address them, forming issue-PR repair pairs.
(2) The data filtering step, which selects and cleans the gathered data to ensure its relevance and accuracy.
(3) The task-oriented data construction step uses the curated data to build datasets specifically designed for different tasks, including Code Question-Answer pairs, Bug Reproduction datasets, Fault Localization datasets, and Code Editing datasets.

\subsection{Raw Data Crawling}
\subsubsection{Repository Selection}
To ensure sufficient diversity within \dataset, we select 30 popular repositories from GitHub, encompassing five widely-used programming languages: Python, Java, JavaScript, TypeScript, and Go. To assure the quality of the issues, we choose repositories that each have over 5,000 stars and are widely utilized within their respective fields.
Examples include Scipy \cite{scipy}, a scientific computing library implemented in Python; Guava \cite{guava}, an open-source utilities library in Java; React \cite{react}, a front-end framework in JavaScript; TypeORM \cite{typeorm}, an object-relational mapping library in TypeScript; and Kubernetes \cite{kubernetes}, a system for managing containerized applications written in Go, and so on.
\subsubsection{GitHub Metadata Crawling}
As the largest platform for hosting and managing software projects, GitHub contains a wealth of metadata. The GitHub REST API is a RESTful programming interface provided by GitHub that allows developers to interact with GitHub services through HTTP requests. Issues and pull requests on GitHub are sequentially numbered, and in the GitHub REST API, each pull request is also considered an issue. We use the GitHub REST API to crawl all issues and pull requests from 30 repositories on GitHub, totaling over 800,000 entries.\footnote{The crawling was performed in August 2024.}
Each entry contains a wealth of metadata, including Base Information, Content Information, Interaction Information and Pull Request-Specific Details. The main features are shown in Table \ref{tab:dataset_info}.

\begin{table}[t]
    \centering
    \caption{Basic information for each entry in GitHub issues and pull requests, including Base Information, Content Information, and Pull Request-Specific Details.}
        \centering
        \footnotesize
        \setlength{\tabcolsep}{1.5mm} 
        \renewcommand{\arraystretch}{1.0}
        \begin{tabular}{c p{9.2cm}} 
            \toprule
             \textbf{Features} &  \textbf{Description} \\ \midrule
            \multicolumn{2}{c}{\textbf{Base Information}} \\ \midrule
            language &  The primary programming language used in the repository.\\
            type &  Whether the entry is an Issue or a PR. \\
            repo &  The repository's name. \\
            owner &  The repository owner, either an individual or an organization. \\
            number &  The unique identifier for the issue or PR. \\
            author-info &  Details about the author of the issue or PR. \\
            author-association &  The author's relationship with the repository (e.g., member).\\
            state &  The current status of the issue or PR (e.g., open, closed). \\
            created-at &  The timestamp when the issue or PR was created. \\
            updated-at &  The timestamp of the most recent update to the issue or PR. \\
            \midrule
            \multicolumn{2}{c}{\textbf{Content Information}} \\ \midrule
            title &  The title of the issue or PR. \\
            body &  The main content or description of the issue or PR. \\
            comments &  The content of comments on the issue or PR. \\
            \midrule
            \multicolumn{2}{c}{\textbf{Pull Request-Specific Details}} \\ \midrule
            merged &  Whether the PR has been merged. \\
            base-info & Information about the base branch of the PR. \\
            head-info & Information about the head branch of the PR. \\
            patch &  The patch details of the PR, describing the changes made. \\
            review-comments & Comments or feedback provided during the PR review process. \\
            \bottomrule
        \end{tabular}
        \label{tab:dataset_info}
    \end{table}

\subsection{Data Pairing}

In addition to the metadata, GitHub issues and PRs often have various interrelationships, connecting information both within and outside the repository. Each issue and PR can be linked to other issues or PRs through specific keywords, status updates, or direct references, creating a complex network of relationships. These connections not only include cross-references between issues and PRs but also extend to links with code commits, branches, tags, and external resources such as other repositories or related documentation. Particularly in the context of problem resolution, PRs typically link directly to the issues they address, establishing a clear fix relationship. This fix relationship provides developers with a transparent path for tracking and resolving issues, and facilitates the integration of automated tools and the overall maintenance of the project.
\subsubsection{Cross Reference Check}
\label{subsub:crossref}

To extract such fix relationships, it is essential to identify both the issue being fixed and the corresponding PR. However, GitHub's REST API does not provide detailed information about cross-reference events directly. It simply provides the names and timestamps of the events associated with the issue or PR, without providing specific information about both sides of the association. To obtain the relationship between a specific issue and the Pull Request that fixes it, we utilize GitHub's GraphQL API, which offers more powerful querying capabilities for retrieving cross-referenced data.

GraphQL is a query language for APIs and a runtime for executing those queries against a type system defined for the data. Unlike RESTful API, which requires multiple requests to gather related data, GraphQL allows for requesting exactly the needed information in a single query. This makes it particularly useful for querying complex relationships, such as those between issues and PRs on GitHub. By leveraging GraphQL, it was possible to efficiently gather the necessary information about linked issues and PRs, enabling a more accurate and comprehensive analysis of fix relationships within the repositories.

With GraphQL, we can identify other issues or PRs that reference a particular issue or PR by querying its CrossReferenceEvent objects. However, there is no direct way to query which issues or PRs the given issue or PR references. To determine the issues or PRs that a specific issue or PR cite, we highlight the process in Algorithm \ref{alg:alg1}.

Algorithm \ref{alg:alg1} takes a GitHub issue or PR entry, as input and returns all other issues and PRs referenced by this entry. The entry is structured as in Table \ref{tab:dataset_info}. In Lines 2-6, the algorithm gathers the content information for each entry listed in Table \ref{tab:dataset_info}, including review comments specifically for PRs. In Line 7, the algorithm parses the collected content to extract all possible references to issues and PRs, including their repository, owner, and number.

According to GitHub's documentation, GitHub designs several rules for automatic linking. The \textit{ExtractCandidates} method identifies potential references to issues and PRs by matching strings with the following patterns:

(1) The full URL of a GitHub issue or PR, which directly includes the repo, owner, and number information of the referenced entry;

(2) Using the "\texttt{\#}" symbol followed by the issue or PR number, such as "\texttt{\#26}";

(3) Using "\texttt{GH-}" or "\texttt{gh-}" (short for GitHub) followed by the issue or PR number, such as "\texttt{GH-26}";

(4) Using the "owner/repo\texttt{\#}" format combined with the issue or PR number, where the owner could be an individual user or an organization on GitHub.

Patterns (1) and (4) explicitly specify the owner and repo, while patterns (2) and (3) only provide the number, which refers to the issue or PR within the current repository.

Lines 8-13 of the Algorithm \ref{alg:alg1} use GraphQL to sequentially query the CrossReferenceEvent for each candidate. For each candidate, the algorithm checks whether it is referenced by the input entry (Line 10). If the candidate is found in the $validReferenceList$, it indicates a confirmed reference, and the candidate is added to the output list (Line 11).

          
    

\begin{algorithm}[t]
    \footnotesize
    \setstretch{0.8}
    \SetAlgoLined
    \footnotesize
    \caption{Reference Extraction for GitHub Issues and PRs}
    \label{alg:alg1}
    \KwIn{$entry$;}
    \KwOut{$referenceList$;}

    $referenceList \gets \emptyset$;

    \uIf{$entry.type == \texttt{"Issue"}$}{
        $content \gets entry.title + entry.body + entry.comments$;}
    \Else{
        $content \gets entry.title + entry.body + entry.comments + entry.reviewComments$;
    }

    \tcp{Extract the numbers, repos, and owners of candidate issues and PRs}
    $candidateList \gets \texttt{ExtractCandidates}(content)$; 

    \ForEach{$candidate \in candidateList$}{
        $validReferenceList \gets \texttt{ExtractInboundReferences}(candidate)$; 

        \If{$entry \in validReferenceList$}{
            $referenceList.\texttt{add}(candidate)$; 
        }
    }
    \Return{$referenceList$;}

\end{algorithm}

\subsubsection{Keyword Check}
After finishing the Cross Reference Check, we obtain lists of references and citations for each issue and PR. Many of these references are not associated with actual fix relationships but are merely mentions. Therefore, we need to filter the extracted references to identify pairs where a PR actually addresses and fixes an issue.
To effectively manage the fix relationships between PRs and issues, GitHub provides nine keywords to link them: \texttt{close}, \texttt{fix}, \texttt{resolve}, and their variations: \texttt{closes}, \texttt{closed}, \texttt{fixes}, \texttt{fixed}, \texttt{resolves}, and \texttt{resolved}.

For PRs targeting the default branch, linking an issue within the same repository can be done by using the rules introduced in Section \ref{subsub:crossref} to represent issues or PRs on GitHub, combined with the appropriate keyword, such as "\texttt{closes} \texttt{\#10}". GitHub will automatically generate a link and display the association on its page. This automated linking assists in tracking and managing code fixes more effectively, ensuring that issues are properly addressed.

It is important to note that we focus on associations within the default branch of the same repository, rather than cross-repository references. The default branch usually represents the main development line, encompassing actual feature updates and bug fixes. By filtering issue and PR pairs in the default branch, we can more accurately identify Fix Pairs, reducing confusion from cross-repository links and enhancing the precision of project management.
For a given PR, we analyze each issue in its reference list that belongs to the same repository. We inspect the content information of these issues to determine if they include the pattern of a keyword followed by an issue identifier. If such a pattern is detected, we classify the issue and PR as a \textbf{fix pair}, indicating that the PR aims to resolve the corresponding issue.

\subsection{Data Filtering}
After completing the Raw Data Crawling and Data Pairing processes, we obtain approximately 500,000 issues and around 100,000 fix pairs. These are intended for constructing the QA dataset and the fault localization and code editing data, respectively. In this section, we describe separate filtering processes for issue and fix pair data. For entries that pass this stage, a semantic check is conducted using both LLMs and human reviewers to select high-quality issues and fix pairs.

\subsubsection{Issue Filtering}
As illustrated in Figure \ref{fig:architecture}, we apply six criteria to filter the issue entries: comment scale, redundancy, external check, event count, robot detection, and participant count. Each criterion is elaborated below:

$\bullet$ \textbf{Comment Scale}: 
We prioritize character count over the number of comments. Entries under 200 characters are removed for lack of content, while those exceeding 10MB or with very high character counts are also excluded to avoid overwhelming the model's capacity. This balances providing sufficient information with staying within the model’s input limits.

$\bullet$ \textbf{Redundancy}: GitHub hosts many duplicate issues, which generally contain repetitive content and lack novelty, making them unsuitable as evaluation datasets. According to GitHub documentation, an issue is marked as a duplicate by GitHub if its comments contain "\texttt{Duplicate of \#}" followed by an issue or PR number. We detect these markers and related keywords to exclude redundant issues, ensuring that each selected issue is unique and valuable for evaluation, thereby enhancing data quality.

$\bullet$ \textbf{External Check}: In reviewing issues, we often encounter a variety of external links, including web pages, images, and videos. Due to the diversity and management challenges of these links, we opt to retain only those issues containing internal GitHub links. This criterion helps avoid the complications of analyzing diverse external content, ensuring our dataset remains focused and manageable, thus improving our data processing and analysis efficiency.

$\bullet$ \textbf{Event Count}: Influenced by methodologies from research like StarCoder\cite{starcoder}, we consider the impact of event numbers within an issue. Issues with more than ten events typically contain a large amount of auto-generated text and robot-produced records, such as extensive logs and unnecessary links. These elements lower data quality and increase processing complexity. Therefore, we exclude issues with more than ten events to maintain a dataset of focused and high-quality entries, enabling more effective analyses.

$\bullet$ \textbf{Robot Detection}: We implemented measures to eliminate robot influences. The initial step involved reviewing the "\texttt{author-info}" field to identify robot users by searching for terms like "bot" in user types and names. Once identified, all issues and responses created by these robots were removed. We also catalogued common robot response patterns to ensure the removal of these automated interactions from the dataset. This thorough approach to robot detection and data cleaning preserves the dataset's integrity and maintains the accuracy of our analyses, ensuring that the insights reflect genuine human interactions.

$\bullet$ \textbf{Participant Count}:
We evaluate the number of participants per issue to ensure discussions in our dataset are lively and diverse. After removing robot users, we exclude issues with only one participant, which often indicate unresolved problems. This approach helps ensure our dataset contains genuine and effective interactions.

\subsubsection{Fix Pair Filtering}
After completing the issue filtering, we thoroughly filter the fix pairs to ensure the data pairs included in the dataset are of high quality and practical value. This filtering process consists of six steps: Merged Check, One-to-One Fix Check, File-Changed Count, Doc Change Detection, Suffix Detection, and Substantial Code Change Detection. Each step is designed to ensure the quality of the fix pairs from different perspectives and to eliminate data that might negatively affect the analysis outcomes. Below, we detail each of these filtering steps.

$\bullet$ \textbf{Merged Check}: We ensure that each PR included has its merged attribute set to true, indicating that the PR has been successfully merged. This confirms that the changes in the PR are accepted and effectively implemented in the project. Unmerged PRs often suggest code issues or non-compliance with project requirements, and including such changes could lead to analytical errors. Thus, this step is crucial for ensuring the quality of the dataset.

$\bullet$ \textbf{One-to-One Fix Check}: We ensure each fix pair corresponds one-to-one by avoiding scenarios where a single PR resolves multiple issues or one issue is addressed by multiple PRs. This step is essential for preventing data coupling and ensuring the clarity and accuracy of our analysis. We scrutinize all fix pairs within the same repository and exclude any that do not meet this strict matching criterion, maintaining a clear structure in the dataset and ensuring accurate reflections of each issue's resolution process.

$\bullet$ \textbf{File-Changed Count}: To simplify task complexity and focus our research, we retain only those PRs involving changes to a single file. This strategy reduces the complexity associated with handling multiple file changes and decreases the likelihood of analysis failure. Additionally, it prevents the invalidation of evaluations due to overly challenging tasks, thus ensuring the practicality and quality of the evaluation dataset.

$\bullet$ \textbf{Doc Change Detection}: To differentiate code changes from documentation updates, we examine the titles of issues and PRs in fix pairs for keywords like ``DOC'', ``Documentation'', or ``READM''. This step helps identify fixes that primarily concern documentation changes. Since our dataset aims to analyze code editing effectiveness, fix pairs focused on documentation are excluded, ensuring our dataset emphasizes code-level changes and enhancing its relevance and validity.

$\bullet$ \textbf{Suffix Detection}: We verify the inclusion of functional code files by checking the suffixes of the modified files to see if they align with the main extensions of common programming languages, such as ``\texttt{.py}'' for Python and ``\texttt{.ts}'' for TypeScript. We also check file paths to exclude any containing the word "test", ensuring that test files are not included in the dataset. This step focuses our attention on changes to operational code, enhancing the dataset’s applicability to real-world scenarios and the precision of our analysis.

$\bullet$ \textbf{Substance Check}: Even when changes involve core files, they may only affect code comments or text strings, which do not impact program functionality. To ensure our dataset includes substantial code changes, we use the tree-sitter tool to analyze changes. This approach helps us exclude fix pairs that solely involve changes to comments or strings, ensuring the dataset contains code changes that could greatly affect program behavior, thereby enhancing dataset quality and ensuring the accuracy and utility of our analyses.

\subsubsection{Semantic Check}
After completing the initial data filtering, conducting a semantic check is crucial to ensure that the selected issues and fix pairs effectively resolve the problems presented. This process aims to further verify and confirm the presentation and resolution of problems at the semantic level, enhancing the practical value and quality of the dataset for research.

(1) \textbf{Automatic Evaluation}: In this step, we utilize the advanced LLM, GPT-4o, to assess whether issues and their corresponding Fix Pairs are structured as clear problem-solution pairs. Specifically, for issue data, GPT-4o targets those presenting a distinct question-and-answer format. The model evaluates the semantic content of the text, focusing on the alignment and efficacy between the problem statements and the proposed solutions. This analysis helps predict the authenticity of the solutions in effectively resolving the issues.

(2) \textbf{Manual Review}: Although LLMs like GPT-4o excel in semantic analysis, there is still a possibility of misjudgments. Therefore, we conduct a manual review of the results evaluated by LLMs to ensure that the selected issues and Fix Pairs indeed propose and resolve problems in their content. This step involves two researchers independently reviewing the selections made by the LLMs, further refining the entries to ensure they effectively address the problems. This approach allows us to validate and supplement the machine’s evaluation from a human perspective, greatly reducing errors and ensuring the quality and usability of the dataset.

By integrating LLM automation with manual review precision, the semantic check process boosts the dataset's reliability and ensures its practical and scientific validity, laying a strong foundation for future research and applications.

\subsection{Task-Oriented Data Construction}
\begin{figure}[t]
	\centering
	\includegraphics[width=1.0\textwidth]{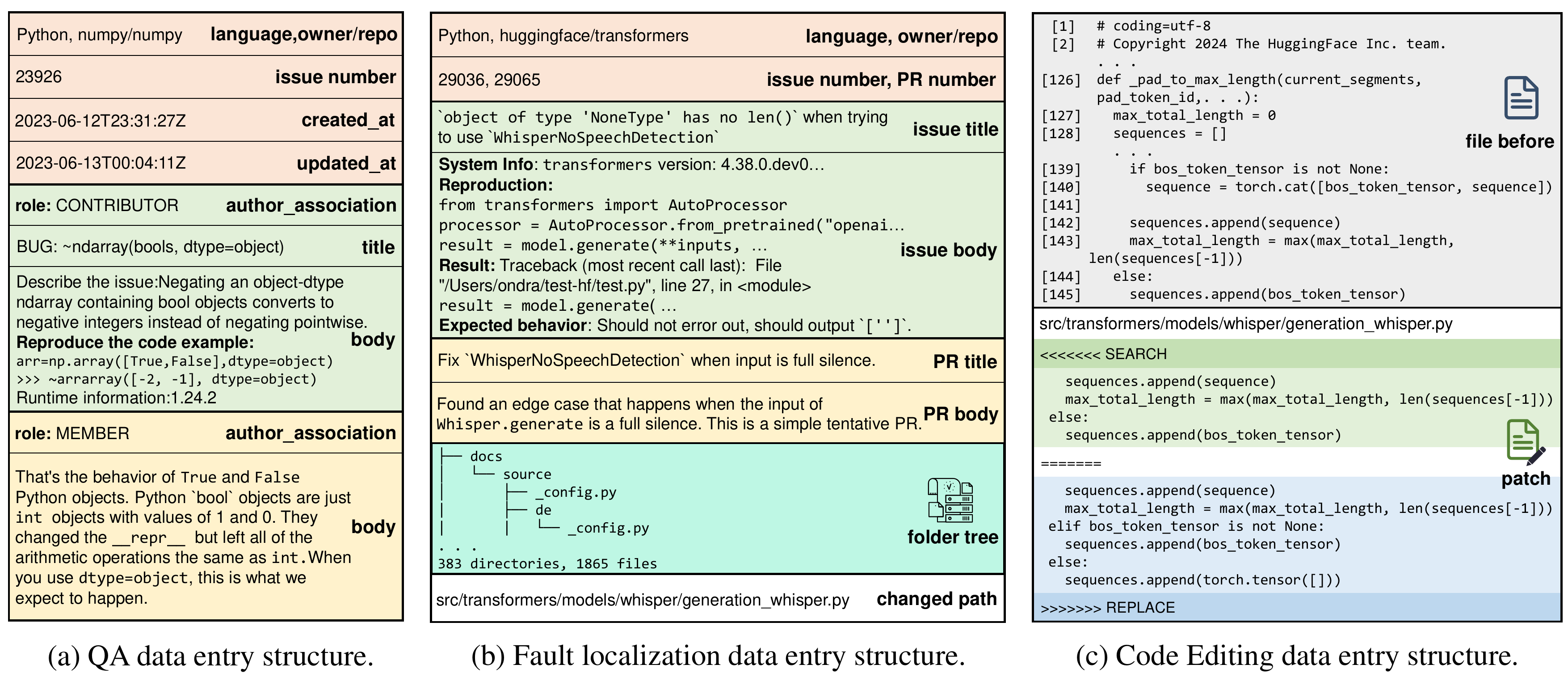}
    \caption{Structures of QA, fault localization, and code editing data entries.}
\label{fig:data_structure}
\end{figure}

This section outlines the creation of specialized datasets for QA, fault localization, and code editing, addressing various research and application needs. These tasks form a logical sequence representing the stages of solving programming challenges: understanding through QA, locating issues via fault localization, and resolving problems through code editing. This structured approach ensures systematic problem-solving supported by robust data and technical validation. The data structures for these tasks are shown in Figure \ref{fig:data_structure}.

\subsubsection{QA Dataset}
To construct the QA dataset, we arrange the data into a question-and-answer format, storing dialogues in a "role-content" format as shown in Figure \ref{fig:data_structure}(a). The "role" indicates the participant's relationship to the repository, defined by the "author-association" field, and "content" includes the dialogue text. The dataset also integrates additional issue-related information such as the repository name, owner, and timestamp of dialogues to enrich the context of each interaction.

The visual representation in Figure \ref{fig:data_structure}(a) details the structure of the dataset where the issue report includes the questioner's author-association, title, and body, while responses consist of the responder’s \texttt{author-association} and the body of the reply. This format facilitates the simulation of realistic question-answer interactions, treating questions as inputs and responses as outputs.

\subsubsection{Fault Localization Dataset}

To construct the Fault Localization dataset, we use the issue description, which includes the issue title and body, along with the folder tree, as inputs. The output is the path of the file that requires change, as depicted in Figure \ref{fig:data_structure} (b). This process begins by downloading the target repository from GitHub and retrieving the state of the repository prior to change using the SHA value provided in the PR’s \texttt{base-info}. Based on this state, we generate a folder structure tree.

To refine the folder structure tree and eliminate extraneous information, we implement pruning operations: we retain only files with common programming language suffixes, delete empty folders, and remove all hidden files and folders that start with a dot (.). These steps focus the dataset on files directly related to programming tasks, ensuring it accurately reflects the locations of files requiring corrections.
Further, Figure 2 (b) illustrates how the issue description along with the folder tree serves as inputs, with the changed path as the output, showing the data flow in the Fault Localization dataset. This structured approach enhances fault localization efficiency and the precision of the correction process.

\subsubsection{Code Editing Dataset}
In constructing the Code Editing dataset, the inputs include descriptions of issues and PRs along with the file before change, encompassing both their titles and bodies. As illustrated in Figure \ref{fig:data_structure} (b) and (c), the issue description and pull description, alongside the file before change, serve as inputs. Typically, the issue serves as the problem statement, while the PR provides the solution. The output is a patch formatted in "SEARCH/REPLACE blocks", which distinctly delineates the code sections requiring change (SEARCH block) and the revised code (REPLACE block).

The advantage of using the "SEARCH/REPLACE blocks" format is its visual clarity in displaying areas of code changes, which facilitates the generation of patches by the model. This format simplifies the model's task of identifying and implementing the necessary code changes.


\section{EXPERIMENTAL SETUP}
For our constructed dataset, \dataset, we propose the following three Research Questions (RQs) based on the specified aspects:
\begin{enumerate}[label=\bfseries RQ\arabic*:,leftmargin=.5in]
    \item How do different models perform in answering questions related to QA, fault localization, and code editing?
    \item What is the impact of each input module in \dataset on the performance of LLMs?
    \item How does the length of an issue affect the LLMs' problem-solving performance?
\end{enumerate}

\definecolor{darkgreen}{rgb}{0,0.5,0}

\begin{table}[t]
    \caption{Selected LLMs, where "-" indicates unknown.}
    \footnotesize
    \renewcommand{\arraystretch}{0.9}
    \centering
    \setlength{\tabcolsep}{1.2mm}
    \begin{tabular}{c|cccccc}
    \toprule
    \textbf{Model} & \textbf{Size} &\textbf{Time} & \textbf{Instruct} & \textbf{Open-source} & \textbf{Max Input Tokens} & \textbf{Max Output Tokens}
\\
\midrule
GPT-4o
& -
& 2024/05
& \cross
& \cross
& 128k
& 16k
\\
GPT-4
& -
& 2023/03
& \cross
& \cross
& 128k
& 4k
\\
Gemini-1.5-Flash
& -
& 2024/05
& \cross
& \cross
& 1m
& 8k
\\
Gemini-1.5-Pro
& -
& 2024/04
& \cross
& \cross
& 1m
& 8k
\\
\midrule
Mistral-large-2
& 123B
& 2024/07
& \tick
& \tick
& 128k
& 4k
\\
CodeQwen-1.5-Chat
& 7B
& 2024/04
& \tick
& \tick
& 64k
& 4k
\\
DeepSeek-Coder-V2
& 236B
& 2024/06
& \tick
& \tick
& 128k
& 4k
\\
DeepSeek-Coder-V2-Lite
& 16B
& 2024/06
& \tick
& \tick
& 64k
& 4k
\\
DeepSeek-Coder
& 33B
& 2023/11
& \tick
& \tick
& 64k
& 4k
\\
DeepSeek-Coder
& 6.7B
& 2023/11
& \tick
& \tick
& 64k
& 4k
\\
    \bottomrule
    \end{tabular}
    \label{tab:models}
\end{table}
\subsection{Model Selection}
As shown in Table \ref{tab:models}, we select ten different large language models (LLMs) to evaluate their performance on coding tasks. These models are broadly used and represent some of the most advanced technologies available. Our selection includes both proprietary models from the GPT and Gemini (GM) series, and other open-source models like Mistral , CodeQwen (CQ) and the DeepSeek Coder (DSC) family, encompassing a range of model sizes. To ensure the validity of all experiments, we strictly control the length of the input data to ensure it does not exceed the maximum input tokens allowed by the selected models.

\subsection{Task Design}
\label{sec:taskdesign}
To comprehensively assess the capability of LLMs in addressing GitHub issues, we designed three distinct tasks: QA, Fault Localization, and Code Editing. We describe the details of each task below.

\textbf{QA:}
The code-related QA task involves enabling the model to answer questions related to code. Specifically, we focus on a single-turn question and answer format, where the title and body of a GitHub issue constitute the issue report, serving as the input. The body of the response from the reply is considered as the reference output. The model needs to understand the issues raised and respond accordingly.

\textbf{Fault Localization:}
The fault localization task aims to identify the file paths that need modification based on the descriptions in GitHub issues. Specifically, the title and body of the issue serve as the issue description, supplemented by the folder tree structure of the repository at that time as inputs. The actual modified file paths, termed as the \texttt{changed path}, serve as the ground truth to validate the accuracy of the model.

\textbf{Code Editing:}
The code editing task involves editing code files according to the descriptions in issues and PRs, as well as the code files before modification. Specifically, the description of the issue and PR (including the title and body) along with the file before editing are used as inputs. The actual changes made to the file, referred to as the patch, serve as the reference output for assessing the performance of the model.

\subsection{Prompt Strategies}
Each task's format consists of a system prompt and a user prompt. The system prompt sets specific requirements and specifies the output format, while the user Prompt displays all the input data the model needs to process.

For the \textit{QA} task, as it involves question and answer, and the output might include code, it is necessary to enclose code blocks within triple backticks (\textasciigrave\textasciigrave\textasciigrave) to ensure the code is correctly formatted. In the \textit{fault localization} task, the output is specifically a file path. To ensure that the LLMs comply with our requirements, we have restricted its output in the prompt to only include the file path. For the \textit{code editing} task, we require the output to be in the form of a "SEARCH/REPLACE" diff file, with specific rules and examples provided. In the user prompt, different data entries should be filled sequentially into the respective fields as described in Section \ref{sec:taskdesign}.

\subsection{Evaluation Metrics}
For tasks involving issue resolution and code fixing, we employ several metrics to evaluate the quality and efficacy of the generated outputs. These include BLEU, ROUGE-L, ROUGE-1, Edit Similarity (ES), Exact Match (EM), and CodeBLEU, and their ranges are all from 0 to 1.


For the QA task, since the output is natural language text, we use BLEU, ROUGE-L, ROUGE-1, and Edit Similarity to evaluate the outputs. For the Fault Localization task, as it outputs only file paths, we solely use EM for evaluation. For the Code Editing task, which involves modifications to code, we exclusively use CodeBLEU, as it is specifically designed to assess code similarity, providing a more accurate measure for this particular task.

\textbf{BLEU} measures the n-gram precision between the generated text and a set of reference texts, incorporating a brevity penalty to discourage overly concise outputs. This metric is primarily used to assess the linguistic accuracy of the generated text.



\textbf{ROUGE} metrics include ROUGE-L and ROUGE-1, which assess the quality of text generated by models in comparison to reference texts. ROUGE-L evaluates sequence-level similarity using the Longest Common Subsequence (LCS) method, focusing on word order and measuring precision and recall by the LCS's proportion to the lengths of the generated and reference texts. The F1 score is emphasized to reflect both structural coherence and content accuracy. ROUGE-1 measures the overlap of single words (1-grams), quantifying precision and recall based on word counts, with the F1 score used as a primary metric to ensure a balanced evaluation of content accuracy and completeness.

\textbf{Edit Similarity} quantifies the similarity based on the minimum number of edits required to transform the generated text into the reference text. This involves calculations of insertions, deletions, and substitutions, providing a detailed look at textual alignment.

\textbf{Exact Match} is a binary metric that scores the output as `1' if it is exactly the same as the reference text, and `0' otherwise. It is used for evaluations where absolute precision in replication of the reference text is required.

\textbf{CodeBLEU} is tailored specifically for code generation tasks, which extends the traditional BLEU by considering aspects pertinent to code such as AST (abstract syntax tree) comparisons, data flow congruence, and naming accuracy.

\subsection{Implementation Details}
In this study, all experiments are conducted on a Linux machine equipped with 256GB of memory. We utilize LLMs configured with a temperature setting of 0.2. For closed-source models, we use the model endpoint API provided by Azure for evaluating GPT models and Google Cloud for Gemini models.
For open-source models, we directly run their released versions from their official repositories according to their documentations, on 8 NVIDIA A100-80G GPUs.

\section{EXPERIMENTAL RESULTS}
In this section, we provide a detailed presentation of our experimental results, including the performance of different LLMs on \dataset.

\begin{table}[!htbp]
    \caption{Performance of various LLMs on \dataset in QA, fault localization, and code editing. In the table, color shades of each block denote performance rankings: darkest for highest, medium for second highest, and lightest for third highest scores.}
    \fontsize{7.5}{11}\selectfont
    \renewcommand{\arraystretch}{0.9}
    \centering
    \setlength{\tabcolsep}{0.4mm}
    \begin{tabular}{c|cccc|cccccc|c}
    \toprule
    \textbf{Metric} & \textbf{GPT-4o} &\textbf{GPT-4} & \textbf{GM-Flash} & \textbf{GM-Pro} & \textbf{Mistral-123B} & \textbf{CQ-7B} & \textbf{DSC-236B} & \textbf{DSC-16B} & \textbf{DSC-33B} & \textbf{DSC-6.7B} & \textbf{Average}
\\
\midrule
\multicolumn{12}{c}{\textbf{QA Evaluation}}
\\
\midrule
BLEU &
0.0943 &
0.1179 &
\cellcolor{black!40}{0.1227} &
\cellcolor{black!20}{0.1208} &
0.0948 &
\cellcolor{black!10}{0.1188} &
0.1102 &
0.0942 &
0.1184 &
0.1182 &
0.1110
\\

ROUGE-L &
0.1189 &
0.1330 &
\cellcolor{black!20}{0.1499} &
\cellcolor{black!40}{0.1551} &
0.1228 &
0.1392 &
0.1340 &
0.1201 &
\cellcolor{black!10}{0.1401} &
0.1366 &
0.1350
\\

ROUGE-1 &
0.1984 &
\cellcolor{black!10}{0.2315} &
\cellcolor{black!20}{0.2410} &
\cellcolor{black!40}{0.2470} &
0.1988 &
0.2264 &
0.2203 &
0.1961 &
0.2272 &
0.2279 &
0.2215
\\

ES &
0.1388 &
0.1715 &
\cellcolor{black!20}{0.1914} &
\cellcolor{black!40}{0.2074} &
0.1411 &
\cellcolor{black!10}{0.1803} &
0.1615 &
0.1406 &
0.1785 &
0.1784 &
0.1689
\\

\hline
Average &
0.1376 &
0.1635 &
\cellcolor{black!20}{0.1762} &
\cellcolor{black!40}{0.1826} &
0.1393 &
\cellcolor{black!10}{0.1662} &
0.1565 &
0.1377 &
0.1660 &
0.1653 &
0.1591
\\
\midrule
\multicolumn{12}{c}{\textbf{Fault Localization Evaluation}}\\
\midrule

EM &
\cellcolor{black!40}{0.6500} &
0.5767 &
0.5900 &
\cellcolor{black!20}{0.6400} &
\cellcolor{black!10}{0.6067} &
0 &
0.5933 &
0.1500 &
0 &
0.0067 &
0.3813
\\

\midrule

\multicolumn{12}{c}{\textbf{Code Editing Evaluation}}\\
\midrule

CodeBLEU &
0.2604 &
0.2447 &
\cellcolor{black!40}{0.3554} &
0.2750 &
0.2889 &
\cellcolor{black!10}{0.3104} &
\cellcolor{black!20}{0.3512} &
0.1969 &
0.2917 &
0.2795 &
0.2854
\\




    \bottomrule
    \end{tabular}
    \label{tab:rq1}
\end{table}
\subsection{RQ1: Performance of LLMs}

To answer RQ1, we evaluate the performance of ten models listed in Table \ref{tab:models}. The results for the evaluation of QA , Fault Localization, and Code Editing tasks are presented in the respective sections of Table \ref{tab:rq1}. We can derive the following observations:

\textbf{The top-performing models differ across the different tasks.}
For the QA task, the models demonstrating the strongest overall performance are Gemini-1.5-Pro, Gemini-1.5-Flash, and CodeQwen-1.5-Chat. In the fault localization task, GPT-4o, Gemini-1.5-Pro, and Mistral-large-2 emerge as the top performers. Lastly, in the code editing domain, the leading models are Gemini-1.5-Flash, DeepSeek-Coder-V2 and CodeQwen-1.5-Chat. A total of six models achieved top-three performance across three tasks, indicating a diverse set of results. This diversity reflects the varying capabilities of different models in addressing distinct aspects of issues. For instance, models like Gemini-1.5-Pro excel in understanding text and handling conversational information; GPT-4o is adept at locating errors based on issue descriptions; and Gemini-1.5-Flash specializes in modifying specified files according to problems and solutions, dealing with specific code. Effectively addressing a particular issue requires a model to possess such atomic capabilities. Our experimental results suggest that the competencies of a model across different dimensions are inconsistent and imbalanced. 
For example, Gemini-1.5-Pro, which ranks among the top two in both QA and Fault Localization, falls into the bottom half in Code Editing.

\textbf{Closed-source models do not always outperform open-source ones, nor do larger models consistently show superior performance.} For instance, GPT-4, a proprietary model, underperforms in fault localization compared to open-source models like Mistral-1.5-large and DeepSeek-Coder-V2, and it does not meet the average performance level in code editing tasks. This demonstrates that proprietary status does not guarantee model superiority. Additionally, the largest and newest model in the DeepSeek series, DeepSeek-Coder-V2, excels in fault localization and code editing but is outperformed by its smaller counterparts, 33B and 6.7B, in QA tasks. This suggests that having more parameters does not automatically translate to better performance, reinforcing that larger models are not always superior across various tasks.

\begin{tcolorbox}
[breakable,width=\linewidth-2pt,boxrule=0pt,top=2pt, bottom=2pt, left=3pt,right=3pt, colback=gray!20,colframe=gray!25]
 \textbf{Finding 1:} The top-performing models vary by task, and it's not always the case that closed-source or larger models outperform their open-source or smaller counterparts.
\end{tcolorbox}



\textbf{LLMs still demonstrate certain limitations in real-world QA scenarios.} Even with Gemini-1.5-Pro, which has the best overall performance, there is still a great discrepancy between its output and the actual feedback. The combined score across four metrics is only 0.1826, with a BLEU score of just 0.1208. Specifically, Gemini-1.5-Pro achieved the highest scores in ROUGE-L and ROUGE-1, with scores of 0.1551 and 0.2470 respectively. This indicates that the vocabulary overlap between the text generated by LLMs and the real-world responses is relatively low, suggesting that the generated content may not effectively capture the key information of the actual responses.

\textbf{Some models do not always follow the specified requirements when executing tasks.} In the Fault Localization task, CodeQwen-1.5-Chat and DeepSeek-Coder 33B recorded EM of zero. Upon examining the output, we found that despite explicitly emphasizing in the prompt,  that only paths should be output, these two models still failed to adhere to the prompt requirements and produced additional explanations instead. This resulted in a failure to achieve an exact match. This also indicates that some models possess limited capabilities in understanding task requirements and are unable to process events according to specified instructions.


\begin{tcolorbox}
[breakable,width=\linewidth-2pt,boxrule=0pt,top=2pt, bottom=2pt, left=3pt,right=3pt, colback=gray!20,colframe=gray!25]
    \textbf{Finding 2:} LLMs exhibit limitations in real-world QA scenarios and some models fail to adhere to specified requirements when executing tasks.
\end{tcolorbox}

\definecolor{deepgreen}{RGB}{0,100,0}
\begin{table}[t]
    \caption{The impact of different module combinations on QA, fault localization and code editing performance. $\Delta$ represents the comparison with the performance when all components are included in the prompt. The average of fault localization is calculated after excluding CodeQwen-1.5-Chat, which scored zero in all cases. IT stands for issue title, IB for issue body, PT for PR title, and PB for PR body.}
    \fontsize{8}{9}\selectfont
    \renewcommand{\arraystretch}{0.9}
    \centering
    \setlength{\tabcolsep}{1.5mm}
    \begin{tabular}{c|c|cc|cc|cc|cc}
    \toprule
    \textbf{Module} & \textbf{Metric} & \textbf{GPT-4o} & \textbf{$\Delta$} & \textbf{DSC-236B} & \textbf{$\Delta$} & \textbf{CQ-7B} & \textbf{$\Delta$} & \textbf{Average} & \textbf{$\Delta$}
\\
\midrule
\multicolumn{10}{c}{\textbf{QA Evaluation}}
\\
\midrule
\multirow{3}{*}{\textbf{BLEU}} &
IT+IB &
0.0943 &
0 &
0.1102 &
0 &
0.1188 &
0 &
0.1078 &
0
\\

&
IT &
0.0843 &
\color{deepgreen}{$\downarrow$10.60\%} &
0.0976 &
\color{deepgreen}{$\downarrow$11.43\%} &
0.0918 &
\color{deepgreen}{$\downarrow$22.73\%} &
0.0912 &
\color{deepgreen}{$\downarrow$15.40\%}
\\

&
IB &
0.0983 &
\color{red}{$\uparrow$4.24\%} &
0.1136 &
\color{red}{$\uparrow$3.09\%} &
0.1190 &
\color{red}{$\uparrow$0.17\%} &
0.1103 &
\color{red}{$\uparrow$2.32\%}
\\

\midrule
\multirow{3}{*}{\textbf{ROUGE-L}} &
IT+IB &
0.1189 &
0 &
0.1340 &
0 &
0.1392 &
0 &
0.1307 &
0
\\

&
IT &
0.0998 &
\color{deepgreen}{$\downarrow$16.06\%} &
0.1095 &
\color{deepgreen}{$\downarrow$18.28\%} &
0.1078 &
\color{deepgreen}{$\downarrow$22.56\%} &
0.1057 &
\color{deepgreen}{$\downarrow$19.13\%}
\\

&
IB &
0.1250 &
\color{red}{$\uparrow$5.13\%} &
0.1399 &
\color{red}{$\uparrow$4.40\%} &
0.1401 &
\color{red}{$\uparrow$0.65\%} &
0.1350 &
\color{red}{$\uparrow$3.29\%}
\\

\midrule

\multirow{3}{*}{\textbf{ROUGE-1}} &
IT+IB &
0.1984 &
0 &
0.2203 &
0 &
0.2264 &
0 &
0.2150 &
0
\\

&
IT &
0.1697 &
\color{deepgreen}{$\downarrow$14.46\%} &
0.1850 & 
\color{deepgreen}{$\downarrow$16.02\%} &
0.1778 &
\color{deepgreen}{$\downarrow$21.47\%} &
0.1775 &
\color{deepgreen}{$\downarrow$17.44\%}
\\

&
IB &
0.2033 &
\color{red}{$\uparrow$2.47\%} &
0.2253 &
\color{red}{$\uparrow$2.27\%} &
0.2276 &
\color{red}{$\uparrow$0.53\%} &
0.2187 &
\color{red}{$\uparrow$1.72\%}
\\

\midrule

\multirow{3}{*}{\textbf{ES}} &
IT+IB &
0.1388 &
0 &
0.1615 &
0 &
0.1803 &
0 &
0.1602 &
0
\\

&
IT &
0.1408 &
\color{red}{$\uparrow$1.44\%} &
0.1575 &
\color{deepgreen}{$\downarrow$2.48\%} &
0.1587 &
\color{deepgreen}{$\downarrow$11.98\%} &
0.1523 &
\color{deepgreen}{$\downarrow$4.93\%}
\\

&
IB &
0.1455 &
\color{red}{$\uparrow$4.83\%} &
0.1680 &
\color{red}{$\uparrow$3.87\%} &
0.1822 &
\color{red}{$\uparrow$1.05\%} &
0.1652 &
\color{red}{$\uparrow$3.12\%}
\\

\midrule

\multirow{3}{*}{\textbf{Average}} &
IT+IB &
0.1376 &
0 &
0.1565 &
0 &
0.1662 &
0 &
0.1534 &
0
\\

&
IT &
0.1237 &
\color{deepgreen}{$\downarrow$10.10\%} &
0.1374 &
\color{deepgreen}{$\downarrow$12.20\%} &
0.1340 &
\color{deepgreen}{$\downarrow$19.37\%} &
0.1317 &
\color{deepgreen}{$\downarrow$14.15\%}
\\

&
IB &
0.1430 &
\color{red}{$\uparrow$3.92\%} &
0.1617 &
\color{red}{$\uparrow$3.32\%} &
0.1672 &
\color{red}{$\uparrow$0.60\%} &
0.1573 &
\color{red}{$\uparrow$2.54\%}
\\

\midrule
\multicolumn{10}{c}{\textbf{Fault Localization Evaluation}}
\\
\midrule

\multirow{3}{*}{\textbf{EM}} &
IT+IB &
0.6500 &
0 &
0.5933 &
0 &
0 &
0 &
0.6217 &
0
\\

&
IT &
0.4533 &
\color{deepgreen}{$\downarrow$30.26\%} &
0.3433 &
\color{deepgreen}{$\downarrow$42.14\%} &
0 &
- &
0.3983 &
\color{deepgreen}{$\downarrow$35.93\%}
\\

&
IB &
0.6400 &
\color{deepgreen}{$\downarrow$1.54\%} &
0.6400 &
\color{red}{$\uparrow$7.87\%} &
0 &
- &
0.6400 &
\color{red}{$\uparrow$2.94\%}
\\

\midrule
\multicolumn{10}{c}{\textbf{Code Editing Evaluation}}
\\
\midrule

\multirow{5}{*}{\textbf{CodeBLEU}} &
IT+IB+PT+PB &
0.2604 &
0 &
0.3512 &
0 &
0.3104 &
0 &
0.3073 &
0
\\

&
IB+PT+PB &
0.2583 &
\color{deepgreen}{$\downarrow$0.81\%} &
0.3525 &
\color{red}{$\uparrow$0.37\%} &
0.3034 &
\color{deepgreen}{$\downarrow$2.26\%} &
0.3047 &
\color{deepgreen}{$\downarrow$0.85\%}
\\

&
IT+PT+PB &
0.2587 &
\color{deepgreen}{$\downarrow$0.65\%} &
0.3255 &
\color{deepgreen}{$\downarrow$7.32\%} &
0.2928 &
\color{deepgreen}{$\downarrow$5.67\%} &
0.2923 &
\color{deepgreen}{$\downarrow$4.88\%}
\\

&
IT+IB+PB &
0.2478 &
\color{deepgreen}{$\downarrow$4.84\%} &
0.3280 &
\color{deepgreen}{$\downarrow$6.61\%} &
0.3005 &
\color{deepgreen}{$\downarrow$3.19\%} &
0.2921 &
\color{deepgreen}{$\downarrow$4.95\%}
\\

&
IT+IB+PT &
0.2425 &
\color{deepgreen}{$\downarrow$6.87\%} &
0.3253 &
\color{deepgreen}{$\downarrow$7.37\%} &
0.2993 &
\color{deepgreen}{$\downarrow$3.58\%} &
0.2890 &
\color{deepgreen}{$\downarrow$5.96\%}
\\

    \bottomrule
    \end{tabular}
    \label{tab:rq2}
\end{table}

\subsection{RQ2: Effectiveness of Different Modules in \dataset Input on LLM Performance} User prompts for different tasks are composed by concatenating various parts of data entries. To explore the impact of each module in issues and PRs on performance, we conduct a series of ablation experiments. We selected one closed-source and two open-source models that performed well in RQ1 for further investigation: GPT-4o, CodeQwen-1.5-Chat, and DeepSeek-Coder-V2. For each experiment, one of these modules is removed from the user prompt, then the model is retested, and the results are recorded in Table \ref{tab:rq2}.

\textbf{The body of a GitHub is crucial for the performance of models in QA and fault localization, while it is not important for code editing.} In the QA task, removing the body from an issue results in a noticeable decline in most metrics, with reductions exceeding 10\%, except for Edit Similarity, which shows minimal change. This indicates that the body of an issue plays a strong guiding role in helping models accurately answer questions, significantly impacting overall performance.
In the fault localization task, removing the body from an issue leads to an average decline of 35.93\% in the Exact Match metric for output paths. Specifically, for DeepSeek-Coder-V2, the Exact Match of predictions fell by 42.14\%, nearly halving. This significant reduction underscores the crucial role that the issue's body plays in aiding the model to accurately locate fault. In the Code Editing task, removing the body of an issue has a minimal impact on CodeBLEU, with the models' average change being only 0.85\%. This indicates that the role of the issue's body varies across different tasks.


\textbf{The title of a GitHub issue may have a negative impact on LLM performance.} Surprisingly, in the QA task, the removal of the issue's title leads to improved performance across all models; similarly, in fault localization tasks, some models like DeepSeek-Coder-V2 show enhanced Exact Match scores after discarding the body. This suggests that not every feature contributes positively to model performance. 
Ideally, the title of a GitHub issue should be a concise summary of the problem by the issue reporter. However, in reality, reporters may not always capture the essence of the issue accurately, leading the title to misdirect the focus. When such titles are placed at the beginning of user prompts, they can potentially misguide the predictions of LLMs, resulting in decreased response quality.


\textbf{The information in a PR plays a certain role in the code editing task.} On average, in the code editing task, removing the title and body of a PR results in a respective decrease of 4.95\% and 5.96\% in the CodeBLEU scores across all models. The title and body of a PR typically contain specific modification suggestions by the author regarding the repository changes. These details are often more direct and explicit in describing the modifications compared to suggestions that might be found in the body of an issue, and they generally do not contain misleading information.

\begin{tcolorbox}
[breakable,width=\linewidth-2pt,boxrule=0pt,top=2pt, bottom=2pt, left=3pt,right=3pt, colback=gray!20,colframe=gray!25]
    \textbf{Finding 3:} In GitHub, the body is crucial for model performance in QA and fault localization tasks, but less important for code editing. Conversely, the title of a GitHub issue can negatively impact LLM performance. Meanwhile, the information in a PR plays a significant role in code editing tasks.
\end{tcolorbox}

\begin{figure}[t]
	\centering
	\includegraphics[width=1.0\textwidth]{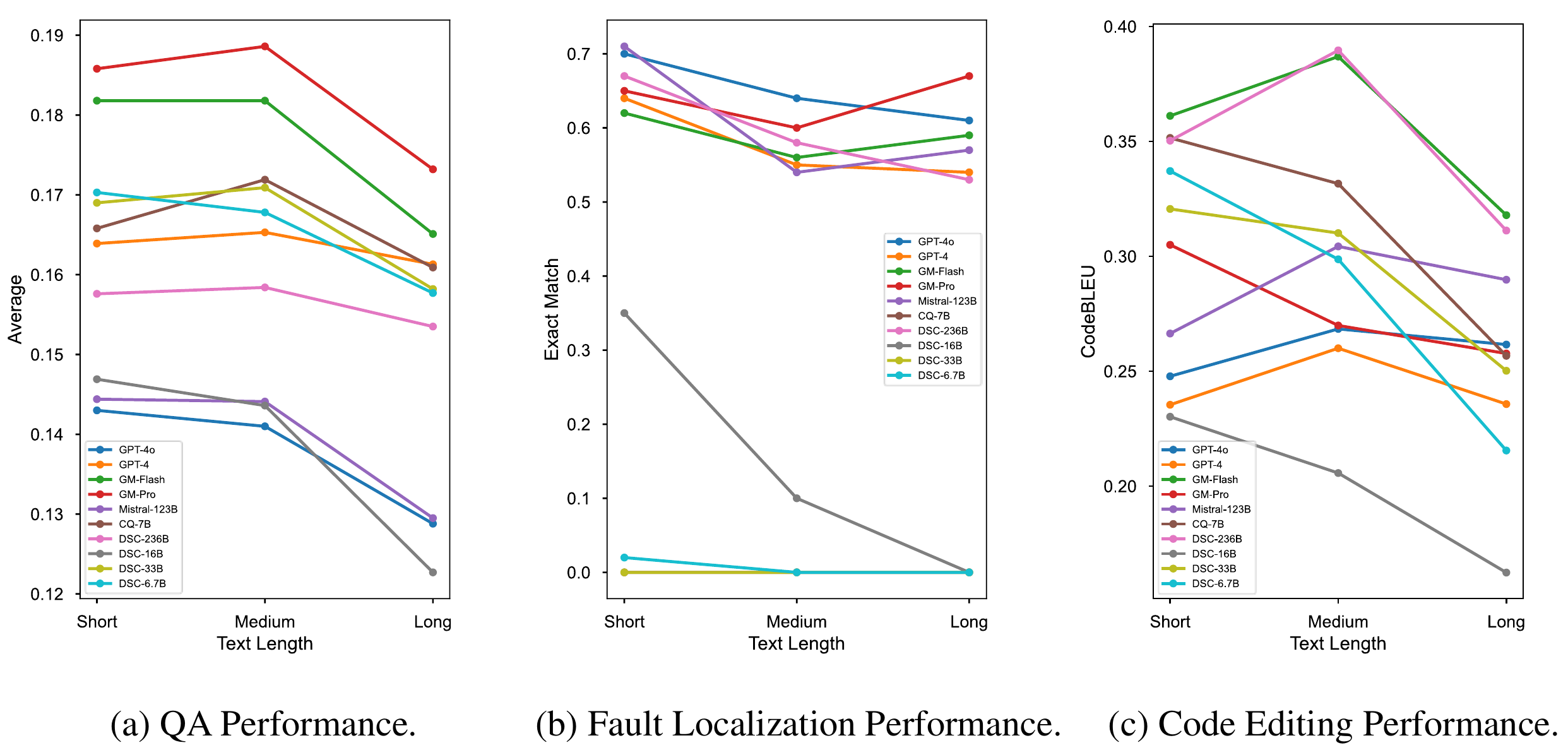}
    \caption{Performance of different LLMs across distinct tasks for short, medium, and long input texts.}
\label{fig:RQ3}
\end{figure}

\subsection{RQ3: Impact of Input Text Length on Issue-Solving Efficacy}
Different large models exhibit varying capabilities when processing information of different lengths. To investigate how issue length affects the performance of LLMs, we sort the dataset by the length of the input text as specified in RQ1 and evenly divide it into three categories: short, medium, and long. We then assess the performance of each group separately across various tasks, with detailed results presented in Figure \ref{fig:RQ3}. We make the following findings:

\textbf{As the length of the input text increases, the performance of each model generally tends to decline.} In QA tasks, it is observed that all LLMs perform the worst with long texts. In Fault Localization, except for two models with EM of zero, other models also show a decline in performance with medium texts compared to short texts. In Code Editing tasks, all models perform worse with long texts than they do with medium texts. Although there are instances where some models perform better with long texts, the overall trend indicates a decline in performance as text length increases. Particularly in the smaller-scale models, DeepSeek-Coder-V2-Lite (16B) and DeepSeek-Coder (6.7B), there is a noticeable decline in performance across various tasks.

\textbf{Different models may excel at processing contexts of varying lengths.} For instance, in the Fault Localization task with medium text lengths, we observe that GPT-4o performs the best, followed by Gemini-1.5-Pro in second place; however, their rankings switch when dealing with long texts. Such ``crossovers'' in rankings occur frequently in Figure \ref{fig:RQ3}, indicating that different models may have varying strengths depending on the text length they process.

\begin{tcolorbox}
[breakable,width=\linewidth-2pt,boxrule=0pt,top=2pt, bottom=2pt, left=3pt,right=3pt, colback=gray!20,colframe=gray!25]
    \textbf{Finding 4:} The performance of models generally tends to decline as the length of the input text increases. However, different models may excel at processing contexts of varying lengths.
\end{tcolorbox}

\section{DISCUSSION}
\subsection{Case Study: Influence of Issue Title}
\label{sec:casestudy}
\begin{figure}[t]
	\centering
	\includegraphics[width=0.8\textwidth]{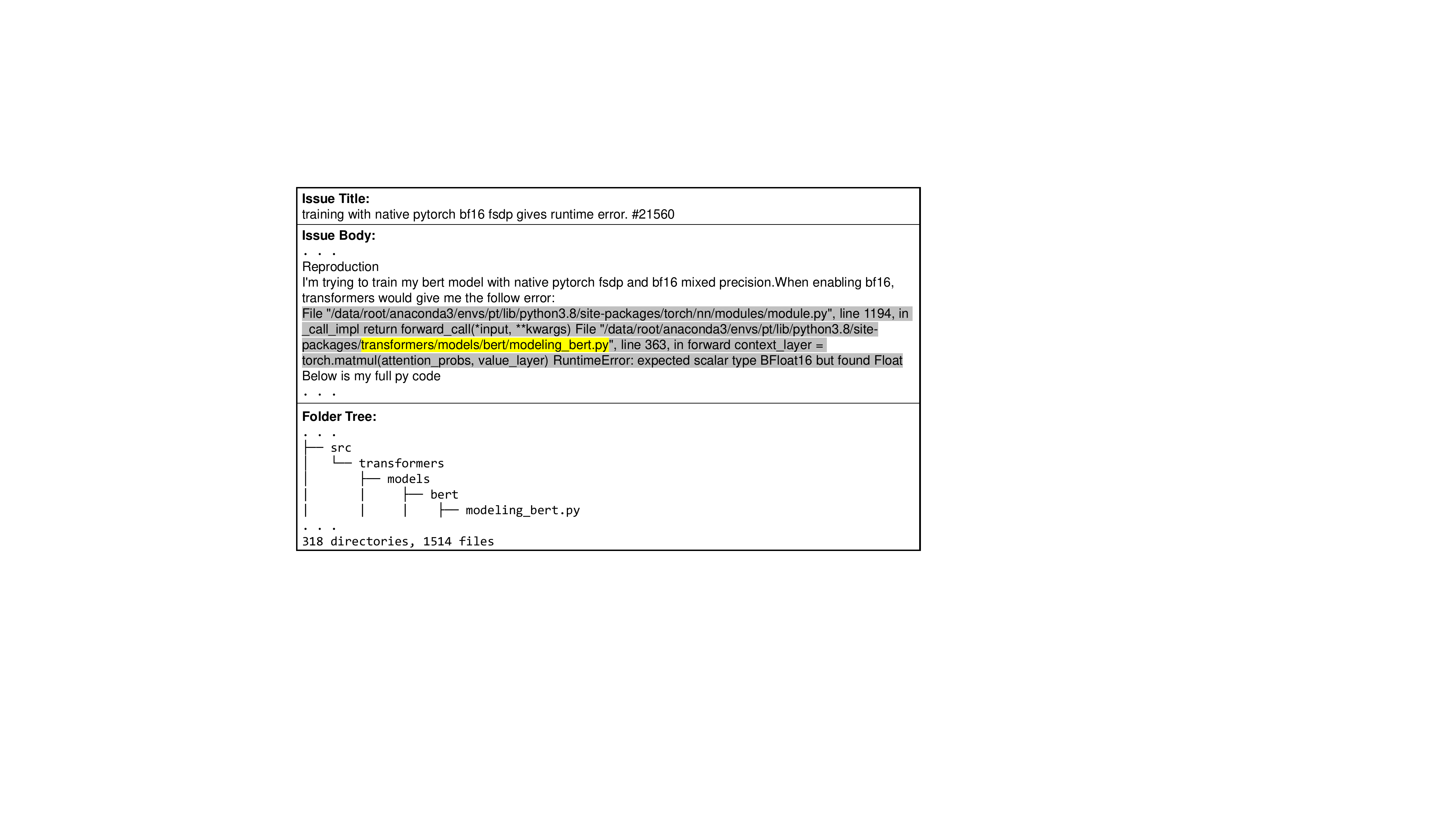}
    \caption{An Example of issue title guiding LLM into the correct response.}
\label{fig:example}
\end{figure}
In RQ2, we discover that the issue body plays a crucial role in QA and fault localization tasks. However, in Table \ref{tab:rq2}, the EM of GPT-4o decreases by 1.54\% under the condition of only retaining the issue body during the fault localization evaluation, which is inconsistent with the overall trend. This indicates that sometimes the issue title can also have a positive effect on predictions.
For example, as shown in Figure \ref{fig:example}, when only the issue body and folder tree are provided to GPT-4o, the model incorrectly locates the error in ``\texttt{src/transformers/models/bert/modeling\_bert.py}'' due to other path information included in the error message. However, when the issue title is added, the model correctly identifies the file that needs modification, ``\texttt{src/transformers/trainer.py}'', guided by the error phenomenon indicated in the title and its understanding of transformers. Therefore, when issue titles are well-formulated, they can provide positive guidance to the model. This also demonstrates that although GitHub issue titles are brief, if well-summarized, they can enable the model to precisely capture the key aspects of the problem.

\subsection{Threats to Validity}

\textbf{Threats to Data Crawling.}
When using the GitHub REST API and GraphQL to crawl issues and PRs, we may encounter network fluctuations and GitHub's rate limiting, which can cause some data downloads to fail. For these failed downloads, we attempted a second download. Despite these efforts, some data still could not be successfully downloaded due to issues like large file sizes or permission restrictions. We tried to address these challenges, but due to time constraints, a small amount of data could not be retrieved. However, this portion of the data often contains inherent flaws and does not impact our subsequent data filtering and analysis.

\textbf{Threats to Language Types and LLMs Selection.}
Due to constraints in the experimental environment and time, \dataset only involves five popular programming languages. We select the most representative repositories within these languages, all of which have well-organized issue management. Additionally, our experimental procedures are easily transferable and replicable for other programming languages. Although we did not explore a broader range of Large Language Models (LLMs) such as Codellama or StarCoder, we choose ten highly representative proprietary and open-source models, with sizes ranging from 6.7B to 236B, to ensure diversity.

\section{CONCLUSION}
In this paper, we introduce \dataset, a novel benchmark for evaluating the atomic capabilities of LLMs in solving real-world GitHub issues, specifically in question-answering, fault localization and code editing. Compared to previous benchmarks used to evaluate the ability of LLMs to resolve issues, \dataset offers a more comprehensive assessment of LLMs' capabilities in addressing various sub-problems. Our experimental results also demonstrate that different models excel at different tasks; moreover, not all information in GitHub issues is helpful in solving problems and may instead misguide the LLMs.

\section*{DATA AVALABILITY}
Our dataset and LLMs' evaluation outputs are available at: \url{https://anonymous.4open.science/status/FAUN-Eval}

\bibliographystyle{ACM-Reference-Format}
\bibliography{references}

\end{document}